# Signatures of Physical Aging and Thixotropy in Aqueous Dispersion of Carbopol

Mayank Agarwal and Yogesh M. Joshi[*]

Department of Chemical Engineering,

Indian Institute of Technology Kanpur, Kanpur 208016, India

* Corresponding Author, E- mail: joshi@iitk.ac.in,

Tel.: +91-512-2597993, Fax: +91-512-2590104

**Abstract**

In this work, we investigate signatures of physical aging in an aqueous dispersion of Carbopol that shows yield stress and weak enhancement in elastic modulus as a function of time. We observe that the creep curves, as well as strain recovery, show a significant dependence on waiting time elapsed since shear melting. The corrected strain, which is the strain in excess of the recovered strain, has been observed to show time – waiting time superposition in the effective time domain, wherein time is normalized by time dependent relaxation time that shows a power-law dependence. The corresponding power law exponent, which is close to unity in a limit of small stresses, decreases with stress and tends to zero as stress approaches the yield stress. For a range of stresses, the material shows time – stress superposition suggesting the shape of the evolving relaxation time spectrum to be independent of the time as well as the stress. This work, therefore, suggests the presence of physical aging in an aqueous dispersion of Carbopol even though the elastic modulus shows only a weak enhancement. We also discuss Andrade type of creep behavior in aqueous Carbopol dispersion.

## I. Introduction

The liquids, depending upon whether their relaxation time is significantly less than or comparable to the observation timescale, respectively show a viscous or viscoelastic rheological response. When relaxation timescale of a material becomes significantly greater than the observation timescale material shows predominantly elastic response.[1] Such materials are usually, but not always, multicomponent mixtures having very high molecular weight species and/or immiscible phases such as particles, bubbles, drops dispersed in a solvent, etc. This type of materials eventually relax completely after a very long time has elapsed. However, over the small observation timescales, they show solid like behavior and demonstrate engineering yield stress, wherein they undergo an extremely slow flow to rapid flow transition at a critical stress.[2-7] Such materials are not intrinsically thixotropic, but owing to large relaxation times, their rheological response may get confused with that of inherently thixotropic response. There exists another class of materials, wherein the crowding or athermal nature of the dispersed ingredients (particles, drops, bubbles, etc.) pose kinetic constraints to obtain the equilibrium structures. Such materials, therefore, are in high free energy state, wherein owing to the mobility of their constituting elements, they explore the phase space so that their structure evolves with time to lower the free energy. This process has been termed as physical aging, which may continue, albeit at a lesser rate, under application of weak stress fields.[8,9] However, the strong stress fields not just stop the physical aging but reverse the same by inducing flow, a behavior known as rejuvenation or shear melting. Such systems may show true yield stress along with thixotropy.[2,10–12] The thermodynamically out of equilibrium materials that have athermal constituents cannot undergo physical aging and therefore remain trapped in high free energy states. These materials show practically infinite viscosity until the applied stress is sufficiently large to break the structure and facilitate yielding.[4,13–15] These materials also can be claimed to show the true yield stress but no noticeable thixotropy.[11,13,14,16] Although the above-mentioned classification appears straightforward, it is not easy to experimentally distinguish the material behavior. In this work, we investigate the rheological behavior of aqueous

Carbopol dispersion that has been proposed to show true yield stress. However, whether this system undergoes physical aging or not has been debated in the literature.[17,18]

Carbopol or Carbomer powder consists of crosslinked polyacrylic acid resin particles,[19–22] and belongs to a broad family of polyelectrolytes.[23] Typically, the particles are polydisperse with a submicron size domain. Upon dispersing the same in water, owing to the acidic nature of the polyacrylic acid, its pH decreases. Usually, acidic dispersion is neutralized with a base, which causes chains to acquire a negative charge that is compensated by counter-ion cations that are present predominantly outside the particle. This causes interconnected links between the crosslinks to repel each other leading to swelling of the particles. It has been reported that the particles absorb water up to 1000 times that of their original volume depending upon nature of copolymer, pH and extent of crosslinking.[19,24–29] A microstructure of a concentrated Carbopol dispersion that shows yield stress has been proposed in the literature.[16,22] According to the same, with an increase in the concentration of Carbopol, the swollen particles cannot occupy the volume without touching each other. Owing to their soft nature, such hindrance from neighboring particles force them to deform in such a fashion that they form flat facets at the contact points, thereby acquiring a polygonal structure.[16,22] In such state individual particle is structurally arrested by the surrounding particles producing a disordered microstructure. Consequently, the dispersion shows a presence of the yield stress, which is usually modeled by the Herschel–Bulkley constitutive equation:[3,4,22,30–32]

$$\sigma = \sigma_y + K\dot{\gamma}^n, \qquad \text{for} \qquad \sigma \geq \sigma_y, \qquad (1)$$

where $\sigma$ is shear stress, $\dot{\gamma}$ is the shear rate, $K$ is consistency and $n$ is the power law index. Interestingly, for a generic class of microgel pastes formed by soft particles, a microstructure very similar to that proposed for concentrated Carbopol dispersion has been suggested in the literature, although with much smaller size (submicron) of the swollen particles.[33,34] Importantly microgel paste has been observed to demonstrate physical aging, wherein owing to the microscopic motion of the soft jammed particles in order to achieve the equilibrium shape, drives the structural recovery.

While it has been established that Carbopol dispersion (beyond a certain concentration, which depends on grade) shows the presence of yield stress, whether this system shows physical aging or not has been discussed in the literature.[17,18] The structural evolution in an out of equilibrium soft material during the quiescent conditions, while reversal of the evolution under application of deformation field renders the same thixotropic character.[2] In a signature test for the thixotropic behavior, a material that has been maintained under quiescent conditions is subjected to up and down shear rate or shear stress ramp. Owing to structural build-up, viscosity shows a higher value during the increasing part of the shear rate/stress. However, rejuvenation at the high stress causes the structure to break, and therefore, viscosity shows a lower value during decreasing part of the cycle leading to hysteresis.[2,35] The literature is full of claims and counterclaims wherein Carbopol dispersion has been observed to demonstrate noticeable hysteresis.[14,17,35–37] On the other hand, several studies suggest the absence of or the presence of negligible hysteresis.[21,38] A comprehensive review of rheological behavior of Carbopol dispersion has been authored by Piau[22] that discusses microstructure and constitutive relations associated with the same.

One of the important characteristic features of the presence of physical aging in a soft material is an observation of aging time-dependent stress overshoot in a step shear rate experiment. Divoux and coworkers[39] observed that for a Carbopol dispersion, under application of moderate magnitudes of shear rates, the intensity of the stress overshoot increases with increase in time elapsed (or aging time) since stopping of shear melting. Furthermore, observation of shear banding in a Couette flow, where stress is either homogeneous or weakly varies in the gap, has also been ascribed to the thixotropic character of a material.[40] Correspondingly there have been reports of transient shear banding in Couette flow of Carbopol dispersion suggesting the presence of physical aging and rejuvenation in the same.[17] On the other hand, some reports do claim absence of shear banding and therefore suggest no noticeable thixotropy.[41,42] Another characteristic feature of thixotropy is breaking of fore-aft symmetry in a flow field around a sphere sedimenting in a fluid. Such behavior is attributed to deformation field dependent aging and rejuvenation that causes viscoelastic properties to be time-dependent. Such breakage of fore-aft symmetry along with negative wake has been reported for Carbopol dispersion by many groups.[43–47]

Lidon and coworkers[18] studied creep and stress relaxation behavior of Carbopol ETD 2050 dispersion. They observe that for creep flow below the yield stress $(\sigma < \sigma_y)$, strain $(\gamma)$ shows a power law dependence on time $(t)$ given by: $\gamma = \gamma_1 + (t/t_0)^\alpha$, where $\gamma_1$, $t_0$, and $\alpha$ are the parameters. They observe that for a broad range of stresses with $\sigma < \sigma_y$, parameter $\alpha$ takes a value around 0.4 that is closer to the value of 0.33 observed by Andrade[48] for metallic wires. Below certain critical stress $\sigma < \sigma_c < \sigma_y$, that depends on the nature of the deformation field during preshear; they report strain to decrease rather than increase during the creep. They attribute this behavior to strain recovery induced by residual stresses when the applied stress is below it. They observe that this strain recovery does not scale with waiting time, and hence propose that rheological response does not suggest any dominant role of physical aging and thixotropy.

More recently, Dinkgreve and coworkers[17] studied the effect of sample preparation protocol on the rheological behavior of Carbopol Ultrez U10 dispersion. They observe the absence of thixotropy for a dispersion obtained through the gentle stirring while noticeable thixotropy for a dispersion prepared using intense, vigorous stirring over a prolonged period. They report that owing to athermal nature of swollen Carbopol particles obtained in the gentle stirring protocol, dispersion does not show time evolution and hence does not show thixotropic behavior. However, during intense stirring, the soft particles break down permanently to form the smaller particles. The Brownian motion associated with the smaller particles create the depletion interactions among the larger particles leading to a gel formation and consequently the thixotropic effects. However, the microgel paste of the soft particles of polyelectrolyte has been observed to show physical aging.[33] The microstructure of the same is very similar to that proposed for the Carbopol dispersion,[16,22] but with much smaller particle sizes (≈230 nm).[33,34] The volume fraction of the soft swollen particles in a microgel suspension significantly exceeds the close packing volume fraction that tends to unity.[33,34,49] Under such circumstances, the individual swollen particles of the microgel are tightly caged by their neighbors leading to the polygonal shapes. Irrespective of very small size of the particles that constitute the microgel paste, Cloitre and coworkers attribute elasticity of the same to elastic contact interactions at the facets claiming it to be athermal in nature.[50] They also model this system by considering the particles to be athermal.[34,50,51] The question, therefore is, whether the mechanism proposed by Dinkgreve and coworkers[17] is generally applicable for all the Carbopol dispersions, or is it specific to the system they investigate.

The literature raises another important question that is how to quantify the thixotropy in soft structurally arrested materials. In a commonly accepted definition of thixotropy proposed by Wagner and Mewis,[2] observation of physical aging and rejuvenation has been considered as a characteristic signature of thixotropy. According to Fielding and coworkers,[52] a system undergoes physical aging when its longest relaxation time evolves with its own age. The presence of hysteresis in an up and down shear rate sweep experiment cannot be used unequivocally to confirm the presence of physical aging/thixotropy as even the equilibrium viscoelastic materials have been observed to show hysteresis in those experiments.[53,54] Furthermore, a non-appearance of transient shear banding also does not imply the absence of physical aging/thixotropy as shown by a recent theoretical study.[40] In addition, there is no enough study to conclusively establish that weak physical aging, wherein relaxation time evolves slower than linear and modulus remains constant with respect to age, leads to breakage of fore-aft symmetry. We, therefore, feel that the most reliable way to quantify physical aging in a material is to rheologically measure relaxation time dependence of waiting time.[8,55–57] In this work, we investigate the possibility of physical aging in an aqueous dispersion of Carbopol 940 microgel using the same methodology. In addition to carrying out the signature test such as step shear rate experiment, we subject the Carbopol dispersion to creep/recovery experiments at different waiting times. The assessment of creep curves at different waiting times and stresses provide vital information about the physical aging behavior of the same. We also discuss the observation of Andrade type of creep behavior in this class of system.

**II. Material, sample preparation, and experimental protocol**

The material used in this work Carbopol 940, present in a white powder form, has been purchased from Loba Chemie Pvt. Ltd, India. The diameter of the particles in the dry powder ranges from 0.1 to 6 μm. Upon dispersing the powder, the water turns into a hazy viscous liquid, wherein individual particles absorb water and swell. In this study, we prepare 0.2 wt. % Carbopol suspension by dispersing Carbopol 940 in ultrapure water (resistivity 18.2 MΩ.cm). In a typical protocol, the dispersion is mixed for one hour by using a magnetic stirrer. After mixing, the dispersion is observed to be a hazy viscous liquid having pH around 5. We use 18 % NaOH solution and adjust the pH of the dispersion to 7 by stirring the sample manually. During the neutralization process, the carboxylic groups in Carbopol ionize and form carboxylate ions in water. These ions repel each other, which augments the flexibility in the polymer to swell and form a gel.[16,21,22,27,31] After neutralization, a sample is kept in ultrasonicator for 30 min to remove the air bubbles, if any, and stored in airtight polypropylene bottles for further analysis. At this stage, the dispersion is observed to be transparent and highly viscous with a paste-like consistency.

In this work, the rheological experiments are performed on TA instruments, AR-G2 rheometer (Couette geometry, bob diameter 28 mm and gap 1.2 mm). We employ serrated geometry with roughness 0.5 micrometer to avoid slippage at the wall. The material stored for around 6 months is loaded into the Couette shear cell at the beginning of each experiment and shear melted under the rotational flow field having a strain rate of 100 s$^{-1}$ for 60 s. Shear melting is necessary to erase the deformation history of the samples that leads to a uniform initial state in all the experiments. We have also checked the consistency and the yield stress of the stored gel at different times within a year, and we found no significant change in the same. In this work, we carry out oscillatory strain/stress sweep (ramp) experiments, oscillatory experiments as a function of time and frequency in the linear viscoelastic domain, shear start-up experiments, creep and strain recovery experiments. In the oscillatory stress sweep experiments, stress amplitude is varied from 0.1 to 100 Pa at a frequency of 1 Hz. This experiment leads to the determination of the linear viscoelastic domain as well as yield stress. After the shear melting, the sample is left to undergo structural evolution (physical aging), if any, for a predetermined time. During this period, an evolution of elastic ($G'$) and viscous ($G''$) moduli are recorded as a function of time at the strain magnitude of 1% and frequency of 1 Hz. The shear start-up experiments are performed at various waiting times by subjecting the samples to a constant shear rate. The creep experiments are carried out at different waiting times after the shear melting is stopped at stresses below and above the yield stress. In all the experiments, over the waiting period, a sample is subjected to oscillatory strain with the magnitude of 1% and the frequency of 1 Hz. Furthermore, the free surface of the suspension is covered with a thin layer of low viscosity silicone oil to prevent evaporation of water. All the experiments are performed at 25°C.

**III. Results and discussion**

After shear melting at different waiting times, the Carbopol is subjected to the oscillatory stress ramp. The corresponding evolution of $G'$ and $G''$ is plotted as a function of the magnitude of stress ($\sigma_0$) in figure 1. It can be seen that $G'$ shows a constant value at lower stresses, and the yield stress ($\sigma_y$) can be observed to be around 20 Pa. We explore the yield stress up to waiting time of 21600 s (= 6 hr) It can be seen that the yield stress of the present system remains constant. With an increase in $\sigma_0$ beyond $\sigma_y$, $G'$ shows a steep decrease. The viscous modulus, $G''$ on the other hand, is significantly smaller than $G'$ for $\sigma_0 < \sigma_y$ and shows a maximum in the vicinity of yielding as commonly observed for various yield stress materials.[58] The system shows a linear response regime for strain magnitude of $\gamma_0 <$ 2 %. In figure 2, we plot $G'$ and $G''$ as a function of the oscillation frequency $\omega$. The elastic modulus $G'$ increases weakly over the explored frequency range of 0.1-100 rad/s with $G' > G''$. The viscous modulus $G''$ shows a minimum, a behavior akin to that observed for concentrated emulsions in a soft glassy state.[59] After ascertaining the linear regime, yield stress, and frequency dependence; we study the aging behavior of the sample by subjecting it to the oscillatory strain field immediately after pre-shear in the linear response regime. In figure 3, we plot the corresponding evolution of $G'$ and $G''$ as a function of time on a logarithmic scale. It can be seen that $G'$ undergoes weak increase while $G''$ shows a weak decrease with time.

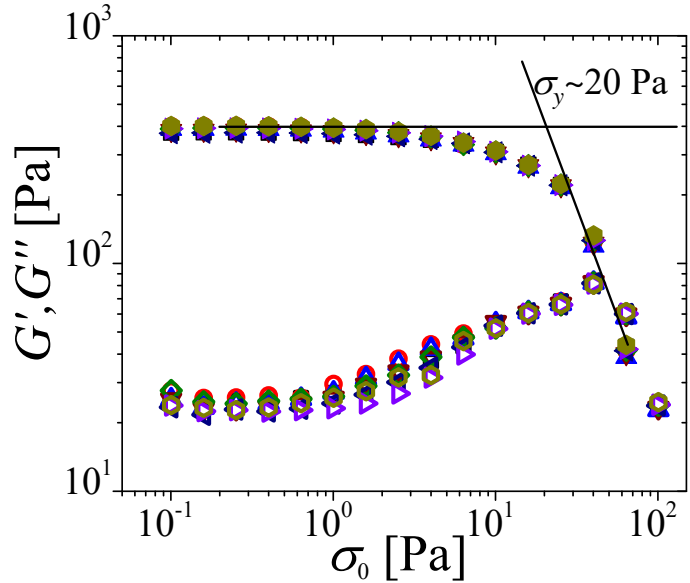

Fig 1. The elastic ($G'$, closed symbols) and viscous ($G''$, open symbols) moduli are plotted with respect to the magnitude of the oscillatory shear stress for 0.2 wt. % Carbopol 940 microgel. The experiment is started at different waiting times: 60 s (square), 300 s (circle), 900 s (up triangle), 1800 s (down triangle ), 3600 s

(diamond), 7200 s (left triangle), 10800 s (right triangle), 21600 s (hexagon) (time elapsed since preshear).

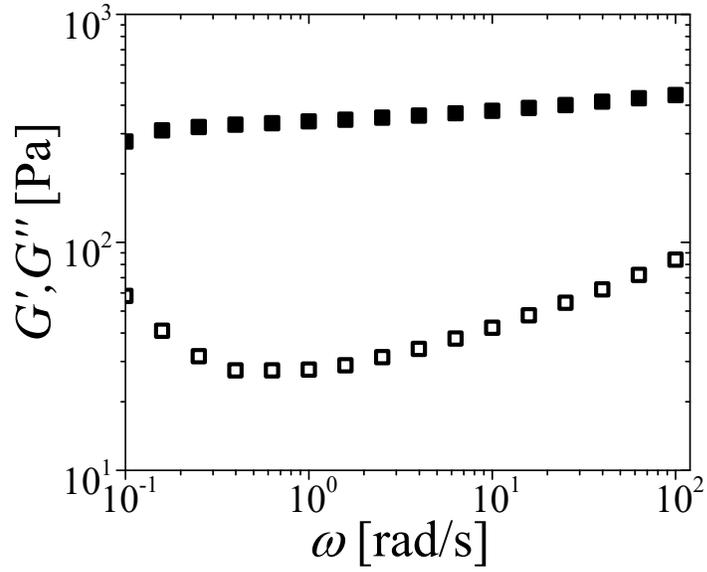

Fig 2. The elastic ($G'$, closed symbols) and viscous ($G''$, open symbols) moduli are plotted against the angular frequency $\omega$ for 0.2 wt. % Carbopol microgel. The experiment is carried out at $\gamma = 1\%$ and waiting time of 600 s (time elapsed since preshear).

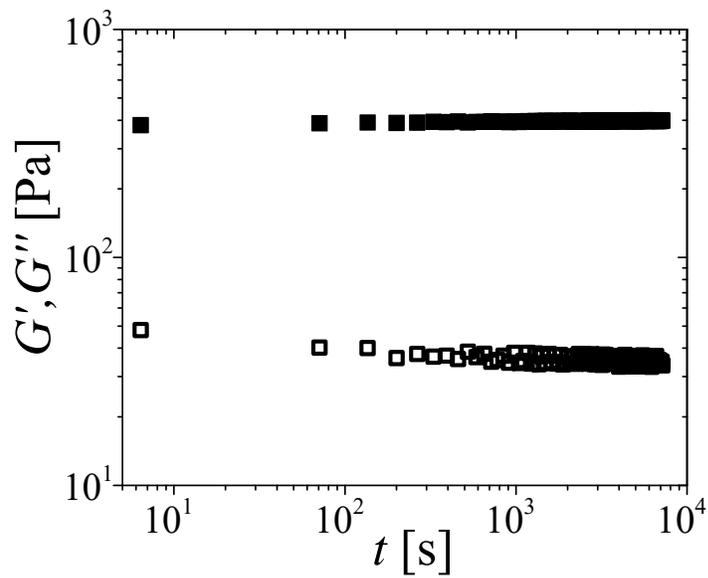

Fig 3. Time evolution of the elastic ($G'$, closed symbols) and viscous ($G''$, open symbols) moduli of 0.2 mass % Carbopol microgel after pre-shear under a small-amplitude oscillatory shear of strain amplitude $\gamma = 1$ % (linear region) and frequency of $\omega = 6.28$ rad/s.

An important characteristic feature of physical aging is the nature of transient viscosity (or stress) overshoot, if any, in a shear start-up experiment.[60-63] In figures 4(a) and (b), we plot transient viscosity $\eta^+$ (given by $\eta^+ = \sigma^+/\dot{\gamma}$) as a function of strain $\gamma$ (given by $\gamma = \dot{\gamma}(t - t_w)$) for a constant $\dot{\gamma}$ applied at time $t_w$ after stopping the pre-shear. In figure 4(a) the data is plotted at different values of applied $\dot{\gamma}$ for $t_w = 60$ s. It can be seen that initially $\eta^+$ increases, reaches a maximum, and shows a very weak decrease with increase in $\gamma$. With an increase in the shear rate, the evolution of viscosity progressively shifts to the lower values suggesting shear thinning nature of the material. In figure 4(b), $\eta^+$ is plotted against $\gamma$ for different waiting times ($t_w$) but for the same $\dot{\gamma}$. The data plotted in the main figure does not suggest any significant difference in the evolution of $\eta^+$ at different $t_w$. However, in the inset of figure 4(b), we show an enlarged region near the maximum in $\eta^+$. Interestingly the data does show an increase in the overshoot, although weak, with an increase in $t_w$. This behavior is suggestive of physical aging occurring during the rest period.[52]

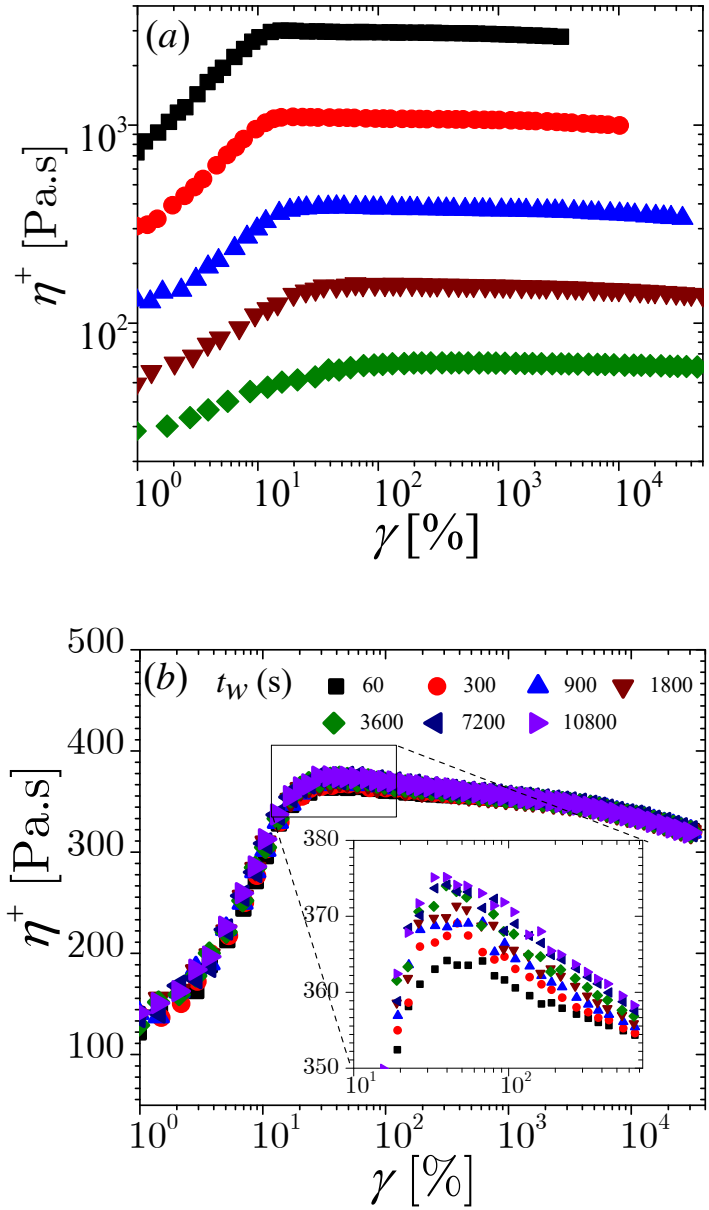

Fig 4. The transient viscosity ($\eta^+$) is plotted as a function of strain ($\gamma$) for different imposed shear rates (figure a, from top to bottom $\dot{\gamma}$ =0.01, 0.03, 0.1, 0.3, 1 s$^{-1}$) at waiting time $t_w = 60$ s. In figure b, the transient viscosity ($\eta^+$) is plotted as a function of strain ($\gamma$) for a step shear rate of $\dot{\gamma}$ =0.1 s$^{-1}$ applied at different waiting times after the pre-shear.

Very weak temporal change in elastic modulus under small amplitude oscillatory shear and observation of a weak increase in an overshoot of $\eta^+$ with aging time qualitatively suggest a possibility of physical aging. Many soft materials, which are thermodynamically out of equilibrium, have been observed to show significant physical aging despite a weak change in elastic modulus as a function of time. To quantitatively assess the physical aging in the Carbopol dispersion, we carry out the creep experiments on the same at different constant stresses at the various waiting times after stopping the pre-shear. During the waiting period we subject the samples to small amplitude oscillatory shear with strain magnitude of 1 % and the frequency of 6.28 rad/s (According to Lidon et al.[18] application of small amplitude oscillatory shear is equivalent to maintaining $\dot{\gamma} = 0$ during the waiting period). In figure 5(a) we plot strain induced in the sample under application of a range of stresses (from 0 Pa to beyond $\sigma_y$) at $t_w = 600$ s as a function of creep time $(t - t_w)$. The strain associated with the stress of 0 Pa merely suggests the strain recovery associated with the sample, whose strain recovery begins at specific $t_w$. We express this strain (recovery) as $\gamma_0$. With an increase in stress, strain increases beyond $\gamma_0$ in such a fashion that at smaller values of stresses (such as for 3 Pa) the strain is still negative, but the recovery decreases compared to $\gamma_0$. At the higher values of stresses (but still lower than $\sigma_y$) strain becomes positive with the modest values. For the stresses greater than $\sigma_y$, however, the strain shows significant upturn suggesting a clear yielding behavior.

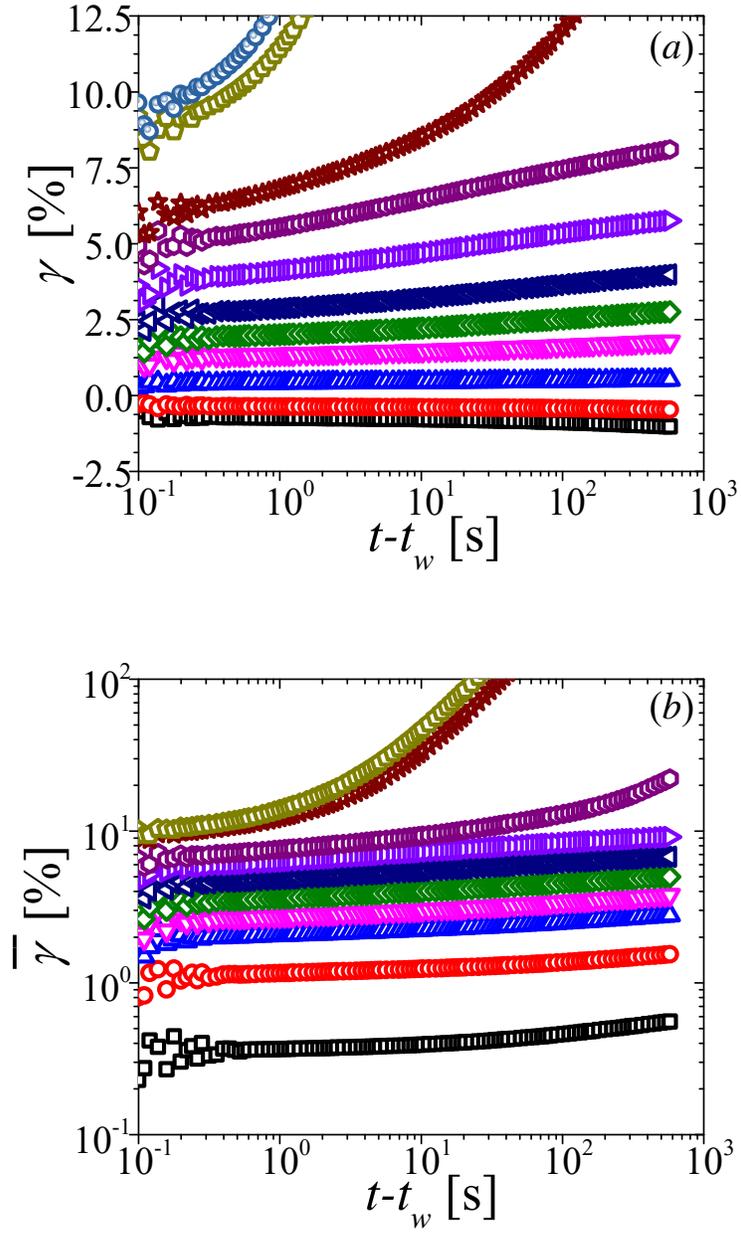

Fig 5 (a) Evolution of strain ($\gamma$) is plotted as a function of creep time under application of constant stresses (from bottom to top 0, 3, 6, 9, 12, 15, 18, 21, 24, 27, 30 Pa) at the waiting time of $t_w = 600$ s. (b) Strain difference $\bar{\gamma} = \gamma - \gamma_0 [\%]$, where $\gamma_0$ is the strain at 0 Pa, is plotted as a function of creep time for the data shown in (a).

For a flow field to which sample is subjected to, the relation between strain rate $\dot{\gamma}$ under application of constant creep stress of $\sigma$ induced in the material can be written as:

$$\sigma = \int_{-t_{w0}}^{0} G(t,t')\dot{\gamma}_{PS}dt' + \int_{0}^{t_w} G(t,t')\gamma_m \omega \cos(\omega t')dt' + \int_{t_w}^{t} G(t,t')\dot{\gamma}dt'$$
$$\phantom{\sigma=}\text{Preshear} \phantom{xxxxxxxxx} \text{SAOS} \phantom{} \text{Creep}$$

(2)

Above equation contains the three terms associated with three steps of application of deformation field: pre-shear at $\dot{\gamma}_{PS}$ applied from $t'=-t_{w0}$ to $t'=0$, followed by SAOS with strain: $\gamma = \gamma_m \sin \omega t$ applied from $t'=0$ to $t'=t_w$ and finally constant creep stress of $\sigma$ applied from $t'=t_w$ to the present time $t'=t$. From the equation, (2) the strain induced in the material for $t>t_w$ is given by $\gamma(t,t_w) = \int_{t_w}^{t} \dot{\gamma} dt'$. For applied stress of $\sigma = 0$, the strain rate associated with the recovery $\dot{\gamma}_0$ can be expressed as:

$$0 = \int_{-t_{w0}}^{0} G(t,t')\dot{\gamma}_{PS}dt' + \int_{0}^{t_w} G(t,t')\gamma_m \omega \cos(\omega t')dt' + \int_{t_w}^{t} G(t,t')\dot{\gamma}_0 dt', \quad (3)$$

wherein the recovered strain in the material for $t>t_w$ is given by $\gamma_0(t,t_w) = \int_{t_w}^{t} \dot{\gamma}_0 dt'$. The strain $\gamma_0$ is due to the recovery of the elastic strain accumulated in the samples. Since during the aging period, we subject the sample to strain controlled oscillatory shear with the magnitude of strain being 1 %, the recovery of the strain begins only at specific $t_w$. In the independent experiments, we observe that this prolonged strain recovery continues over the explored time-scale of several hundreds of seconds as also observed by Lidon and coworkers.[18] When the sample gets subjected to non-zero stresses, the strain induced in the sample is over and above $\gamma_0$. Since the Boltzmann superposition principle of linear viscoelasticity suggests that for the independent application of deformation fields the response is linearly additive, the strain response to the applied non-zero stress can simply be estimated by subtracting $\gamma_0$ from the respective strain $\gamma$. Therefore, subtracting equation (2) from equation (3) leads to:

$$\sigma(t) = \int_{t_w}^{t} G(t,t')(\dot{\gamma} - \dot{\gamma}_0)dt' = \int_{t_w}^{t} G(t,t')\overline{\dot{\gamma}}dt', \tag{4}$$

where $\overline{\dot{\gamma}} = \dot{\gamma} - \dot{\gamma}_0$ and $\overline{\gamma}(t,t_w) = \int_{t_w}^{t}(\dot{\gamma} - \dot{\gamma}_0)dt'$. Here, we acknowledge that deformation field during the shear melting step is non-linear, however, at the end of shear melting that is at $t' = 0$, since the state of the sample is identical irrespective of an experiment, subtraction of equation (3) from equation (2) identically leads to the equation (4). In figure 5(b), we plot $\overline{\gamma}(=\gamma - \gamma_0)$ as a function of $t - t_w$ for the same data shown in figure 5(a). As expected, for all the stresses $\sigma < \sigma_y$, $\overline{\gamma}$ is positive but shows a very slow increase. For $\sigma > \sigma_y$, on the other hand, $\overline{\gamma}$ show a rapid increase. The system, therefore, clearly shows yielding through viscosity bifurcation for stresses in the vicinity of $\sigma_y$.

In figure 6, we explicitly demonstrate the process of how $\gamma_0$ is subtracted from $\gamma$ to obtain $\overline{\gamma}$ for two very different stresses (3 Pa and 15 Pa) at two extreme waiting times ($t_w$ =120 s and 600 s). As expected, the magnitude of $\gamma_0$ (associated with $\sigma = 0$) decreases with increase in $t_w$. On the other hand, for $\sigma = 3$ Pa strain evolves in a negative direction but the magnitude of $\gamma$ increases with increase in $t_w$. However, as discussed before, it is misleading to consider $\gamma$ to be an actual strain as it is over and above $\gamma_0$. Consequently, $\overline{\gamma}(=\gamma - \gamma_0)$, plotted as a function of $t - t_w$ for $\sigma = 3$ Pa, shows an expected decrease with an increase in $t_w$. Contrary to that observed for $\sigma = 3$ Pa, for $\sigma = 15$ Pa, $\gamma$ is not just positive but its magnitude decreases with increase in $t_w$. As expected, the corresponding $\overline{\gamma}$ also shows the same trend for $\sigma = 15$ Pa as shown for $\sigma = 3$ Pa. The creep data in a limit of small times $t - t_w \rightarrow 0$ (in the present case this limit is, as soon as the inertial strain oscillations attenuate) also leads to instantaneous elastic modulus associated with the material. If we calculate modulus based on $\overline{\gamma}$ given by: $G = \sigma/\overline{\gamma}$, for $t_w$ =120 s, we get $G \approx$ 393 Pa for $\sigma = 3$ Pa while $G \approx 395$ Pa for $\sigma = 15$ Pa. The former value matches very well with the value of $G' \approx 391$ Pa obtained in SAOS experiments at $t_w = 120$ s as shown in figure 3. For $t_w = 600$ s the modulus values based on $\bar{\gamma}$ is $G \approx 826$ Pa for $\sigma = 3$ Pa and $G \approx 427$ Pa for $\sigma = 15$ Pa, whereas at $t_w = 600$ s, $G'$ shows only a small increase with the value of 395 Pa as plotted in figure 3. The actual value of uncorrected strain ($\gamma$) at $\sigma = 15$ Pa is smaller than $\bar{\gamma}$, consequently modulus based on $\gamma$ is expected to be larger than that of based on $\bar{\gamma}$, and therefore significantly larger than $G'$ at respective $t_w$. The above discussion suggests that while $G'$ shows only a marginal increase in SAOS experiment, in the creep experiments increase in modulus based on $\bar{\gamma}$ is larger due to correction associated with the strain recovery.

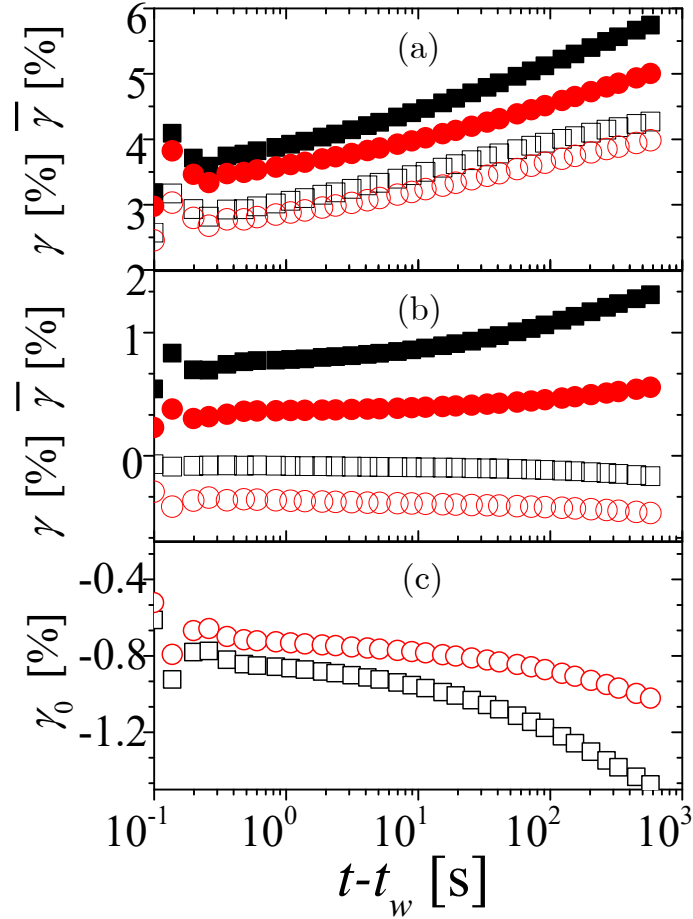

Fig 6. Actual strain ($\gamma$, open symbols) induced in the material is plotted as a function of $t - t_w$ for $t_w = 120$ s (square) and 600 s (circle) for constant creep stress

of 15 Pa (a), 3 Pa (b) and 0 Pa (c). The strain associated with 0 Pa stress has been termed as $\gamma_0$. In figures (a) and (b) we also plot the strain difference $\bar{\gamma} = \gamma - \gamma_0$ [%] (filled symbols) as a function of $t - t_w$.

In the inset of figure 7, we plot $\bar{\gamma}$ as a function of $t - t_w$ for the creep experiments with $\sigma = 3$ Pa started at different $t_w$. It can be seen that $\bar{\gamma}$ induced in the system depends not just on creep time $(t - t_w)$ but also on the time $t_w$ at which the creep experiments are started (the same behavior can also be seen for $\sigma = 15$ Pa as shown in figure 6). In case of equilibrium viscoelastic materials that do not undergo physical aging but do show strain recovery under zero stress, $\bar{\gamma}$ at any $t_w$ is only a function of $t - t_w$. (We do confirm this for an ergodic viscoelastic material: 0.5 weight % aqueous Polyacrylamide (PAm) solution (HiMedia, Mol. Wt. 50,00,000) under application of identical deformation field employed in this work on the same rheometer. For aqueous PAm solution while the strain recovery shows behavior similar to that shown in figure 6(c) and $\gamma$ versus $t - t_w$ does depend on $t_w$, the corrected strain $\bar{\gamma}$ is only a function of $t - t_w$ and is independent of $t_w$). Furthermore, the present material shows that $\bar{\gamma}$ plotted against $t - t_w$ decreases with increase in $t_w$. This behavior suggests that the sample undergoes structural evolution or physical aging as a function of time, which makes it stiffer with an increase in $t_w$ causing lesser $\bar{\gamma}$ to get induced in the same.

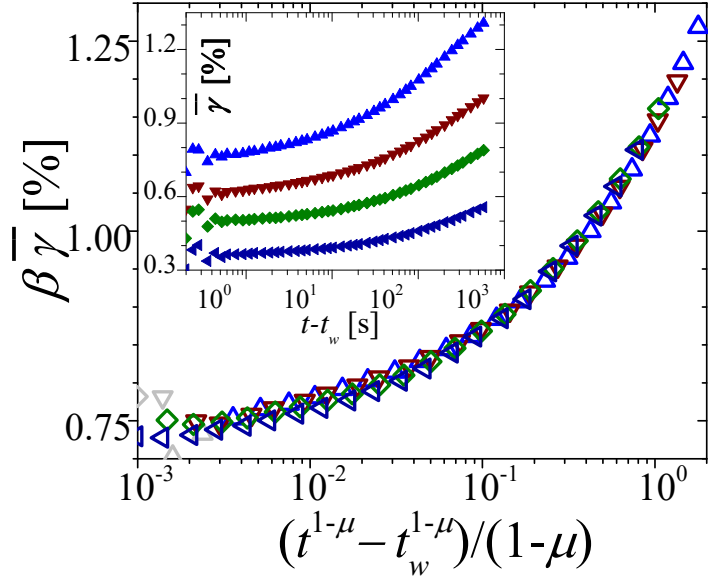

Fig 7. In the inset, the strain difference $\bar{\gamma} = \gamma - \gamma_0$ [%] is plotted as a function of creep time for a constant creep stress of 3 Pa for different $t_w$ (from top to bottom: $t_w$ = 120 s (up triangle), 240 s (down triangle), 400 s (diamond), 600 s (left triangle)). Main figure 7 shows the corresponding time – aging time superposition as discussed in the text.

The observed creep behavior that shows the signatures of physical aging does not validate the time-translational invariance that necessitates material properties to be independent of time, and therefore equation (4) can be modified to:[52]

$$\sigma(t) = \int_{t_w}^{t} G(t-t', t') \dot{\bar{\gamma}} dt'. \tag{5}$$

While undergoing physical aging, if a material preserves the shape of the spectrum of relaxation times, there always exists an effective time $(\xi(t))$ that depends on real-time defined as:[8,57]

$$d\xi/\tau_0 = dt'/\tau(t') \quad \Leftrightarrow \quad \xi(t) = \tau_0 \int_0^t dt'/\tau(t'), \tag{6}$$

that transforms the equation (5) to: [64,65]

$$\sigma[\xi(t)] = \int_{\xi_w}^{\xi} G(\xi - \xi') \frac{d\overline{\gamma}}{d\xi'} d\xi'. \tag{7}$$

In equations (6) and (7), $\tau = \tau(t')$ represents the dependence of mean relaxation time on time, $\xi_w = \xi(t_w)$ and $\xi' = \xi(t')$. In the effective time domain material possesses constant relaxation time $\tau_0$, whose actual value does not influence the analysis.[64,65] Equation (7) suggests that material validates *effective* time-translational invariance. Furthermore, the existence of the Boltzmann superposition principle in the effective time domain with strain as an independent variable always validates the existence of the same with stress as an independent variable given by:[57,65]

$$\overline{\gamma}[\xi(t)] = \int_{\xi_w}^{\xi} J(\xi - \xi') \frac{d\sigma}{d\xi'} d\xi', \tag{8}$$

where $J(\xi - \xi')$ is the creep compliance.

The definition of the effective time domain given by equation (6) requires prior knowledge of the dependence of mean relaxation time on time: $\tau = \tau(t)$.[65,66] To obtain $\tau = \tau(t)$ for the Carbopol dispersion studied in this work, we carry out the horizontal and vertical shifting of the $\overline{\gamma}$ data shown in the inset of figure 7 in a limit of $t - t_w \ll t_w$. As discussed in greater detail elsewhere,[8,66,67] the corresponding horizontal shift factor suggests that the relaxation time shows a power law dependence on time given by $\tau \propto t^\mu$. This dependence in a dimensionally consistent form can be written as[52,57,68]:

$$\tau = A\tau_m^{1-\mu} t^\mu, \tag{9}$$

where $A$ is a constant, $\tau_m$ is the characteristic relaxation time associated with the aging process and $\mu$ is the power law coefficient. Interestingly the power law dependence given by equation (9) has been observed for a variety of non-ergodic systems such as polymeric glasses,[8,9] colloidal glasses,[56,69] soft glassy materials,[55,57,67,70] spin glasses,[71] etc. Incorporation of the equation (9) into (6) and assigning $\tau_0 = \tau_m$ leads to:

$$\xi - \xi_w = \tau_0 \int_{t_w}^{t} \frac{dt'}{\tau(t')} = \frac{\tau_0^\mu}{A}\left[\frac{t^{1-\mu} - t_w^{1-\mu}}{1-\mu}\right].$$

(10)

In figure 7, we plot vertically shifted $\bar{\gamma}$ as a function of $\left(t^{1-\mu} - t_w^{1-\mu}\right)/(1-\mu)$ for the creep data associated with $\sigma = 3$ Pa. It can be seen that aging time-dependent $\bar{\gamma}$ versus $t - t_w$ data shown in the inset of figure 7 demonstrates a remarkable superposition in the effective domain.

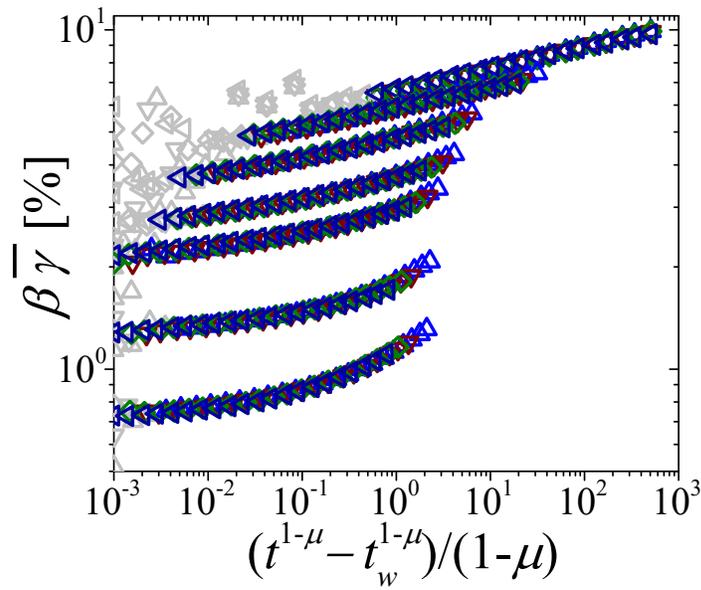

Fig 8. Time-aging time superpositions of vertically shifted $\bar{\gamma}$ at different constant stresses (from bottom to top 3, 6, 9, 12, 15, 18, 21 Pa) are plotted in the effective time domain.

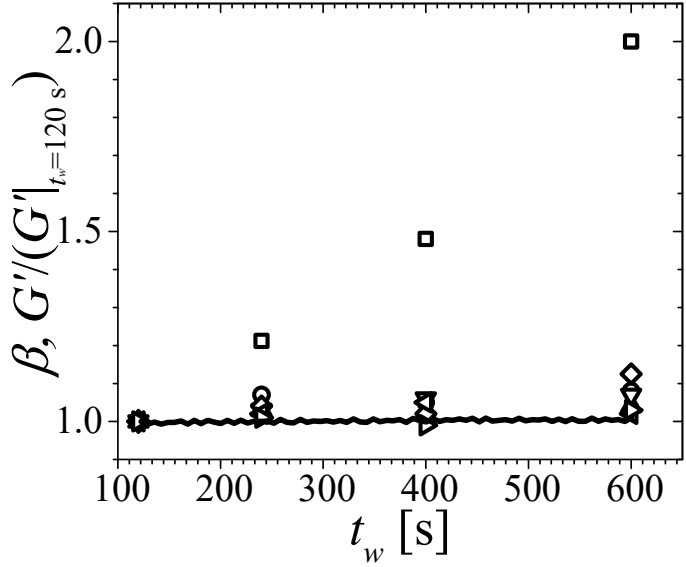

Fig 9. Vertical shift factor $\beta$ is plotted as a function of $t_w$ waiting time for different stresses (3 (square), 6 (circle), 9 (up triangle), 12 (down triangle), 15 (diamond), 18 (left triangle), 21 Pa (right triangle)). We also plot elastic modulus normalized by that of at 120 s, $G'/G'\big|_{t_w=120\text{ s}}$ (line).

We carry out this analysis for the creep experiments performed at all the explored stresses below and above the yield stress. The corresponding superpositions, wherein vertically shifted $\bar{\gamma}$ is plotted against $\left(t^{1-\mu} - t_w^{1-\mu}\right)/(1-\mu)$, are shown in figure 8. It can be seen that $\bar{\gamma}$ shows good superpositions for all the explored stresses below and above the yield stress. In figure 9, we plot the vertical shift factor $\beta$ as a function of $t_w$ for all the explored stresses. As derived by Kaushal and Joshi[65] while formally developing the effective time domain theory from the generalized Maxwell and Voigt model, the vertical shift factor $\beta$ is related to change in modulus as a function of time. It can be seen that, except for $\sigma = 3$ Pa, $\beta$ increases just slightly with $t_w$. For $\sigma = 3$ Pa, on the other hand, $\beta$ shows a significant increase with $t_w$. Such behavior of $\beta$ is also apparent from figure 6, wherein change in initial $\bar{\gamma}$ with $t_w$ is significant for $\sigma = 3$ Pa than at higher stresses. Since reference waiting time is $t_w$ =120 s (for which $\beta = 1$) and $\beta$ is a vertical shift factor associated with the compliance, we also plot elastic modulus normalized by that of at 120 s, $G'/G'|_{t_w=120\text{ s}}$. It can be seen that over the explored waiting times, there is hardly any change in modulus leading to: $G'/G'|_{t_w=120\text{ s}} \approx 1$. This suggests that correction of strain by that of recovered strain indeed does lead to a different value of modulus scale compared to what is obtained by measurement of $G'$. Furthermore, this difference is very significant for $\sigma$ =3 Pa, for which the strain response was observed to be negative than at the higher stresses. In addition, the value of modulus is also expected to be dependent on the applied creep stress.

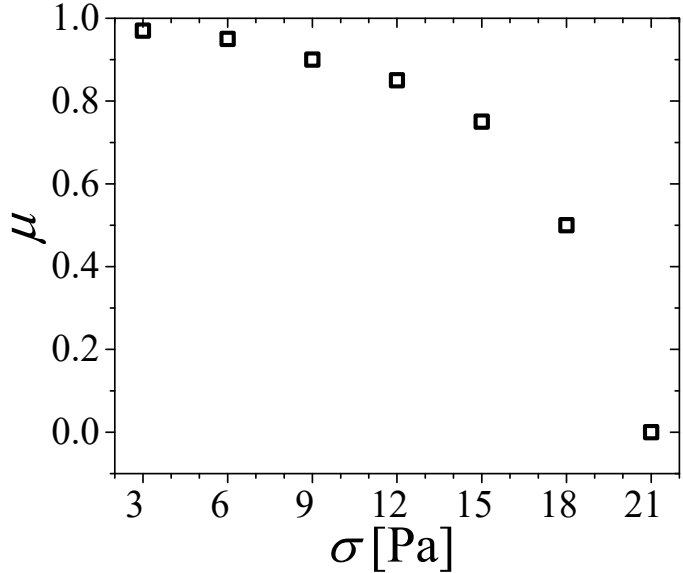

Fig 10. Power law exponent $\mu$ needed to obtain the superposition is plotted as a function of creep stress.

In figure 10, we plot the power law coefficient $\mu$ as mentioned in equations (9) and (10) as a function of $\sigma$. It can be seen that $\mu$ decreases with an increase in $\sigma$, and in a limit of $\sigma \to \sigma_y$ the system ceases to show any time dependence leading to $\mu \to 0$. For $\sigma > \sigma_y$, the material yields erasing all the aging history and therefore creep curves superpose without any shifting. This behavior is very similar to that observed for microgel paste that undergoes physical aging and has similar microstructure to that of Carbopol dispersion.[55] Closer observation of the superpositions up to $\sigma = 9\,\text{Pa}$ suggests that the superpositions show self-similar curvature. As a result, horizontal and vertical shifting of the same leads to time – aging time – stress superposition as shown in figure 11. The corresponding shift factors are plotted in the inset of figure 11. The time – aging time superpositions for stresses greater than 9 Pa and closer to the yield stress, however, show noticeably different curvature and therefore do not participate in the time – aging time – stress superposition. It should be noted that curvature of the creep compliance leads the retardation time spectrum in the equilibrium materials.[1] The curvature of the creep compliance in the effective time domain, on the other hand, leads to an instantaneous shape of the retardation time spectrum in a time-dependent material. The shape of the retardation time spectrum always fixes the shape of the relaxation time spectrum.[1] The very fact that $\bar{\gamma}$ shows superposition in the effective time domain, therefore, suggests that under application of a given constant value of stress, the mean relaxation time increases with $t_w$ while preserving the shape of the relaxation time spectrum. The observation of time – aging time – stress superposition, then again, suggests that the domain of stresses over which such superposition is observed, the shape of relaxation time spectrum remains unaltered with $t_w$ as well as $\sigma$, though the mean relaxation time may change with both the variables.[68,72] However, as the stress gets closer to the yield stress, the superpositions do not participate in the time – aging time – stress superposition. This behavior indicates alteration of the shape of the relaxation time spectrum as well as its mean value with $\sigma$. The waiting (aging) time dependence of the creep data shown by the creep curves, the observation of time – aging time – stress superposition shown by the same, decrease in $\mu$ as a function of $\sigma$, and cessation of time dependency $(\mu = 0)$ above $\sigma_y$ are all the standard characteristic features of physical aging reported for various kinds of out of equilibrium soft materials.[4,55,68,73–76]

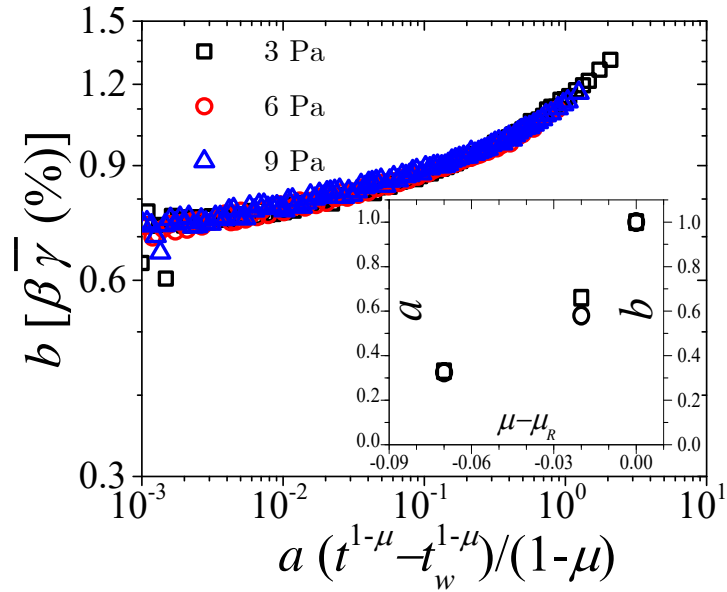

Fig 11. Time-aging time-stress superpositions for the data shown in fig 8 are plotted in the effective time domain. Inset: The horizontal shift factor ($a$) (square) and vertical shift factor ($b$) (circle) is plotted as a function of $\mu - \mu_R$. where $\mu_R = 0.97$ is the value of $\mu$ at reference stress 3 Pa.

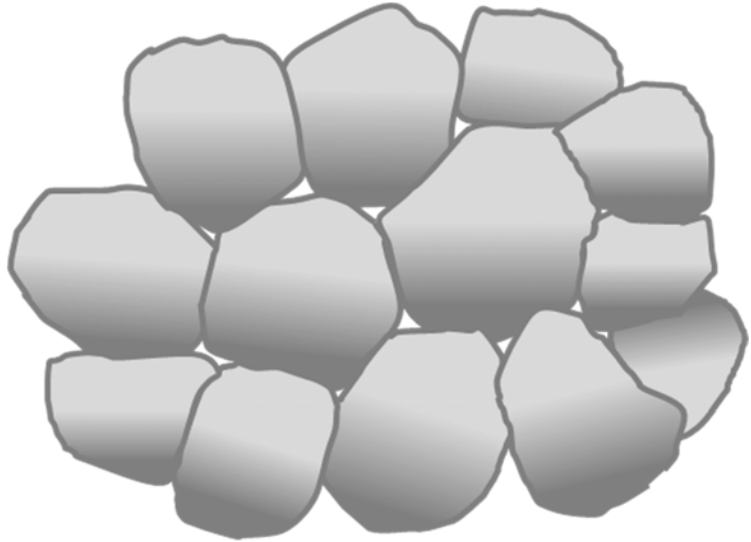

Fig 12. Schematic representing proposed microstructure of the Carbopol dispersion microgel in a swollen state

We shall now discuss the possible origin of physical aging in an aqueous Carbopol dispersion studied in this work. In figure 12, we schematically represent a proposed microstructure associated with the Carbopol 940 dispersion that closely follows the same mentioned in the literature for Carbopol dispersion[16,22] as well as the microgel paste.[33,55] In the Carbopol dispersion, since the volume fraction of the swollen (soft) particles approaches unity, owing to steric constraints, the particles acquire polygonal shape leading to flat facets at contacts. The shape and orientation of the individual particles is dictated by their local neighbors leading to a disordered metastable microstructure. When such material is shear melted at high stresses, the soft particles reorient and flow past each other owing to thin water film lubricating the flat facets of the same.[55] During the process of fluidization, the system forgets any previous shear history. After the shear melting is stopped, the soft particles get trapped in an arrangement associated with the fluidized state, which intrinsically is a high free energy state. Such a state is an outcome of excess deformation of the swollen particles from their respective equilibrium shapes. As a result, the stress gets induced in the same. This internal stress or deformation of shape, away from that of equilibrium induces structural rearrangement when subjected to quiescent conditions. Consequently, the individual particles in a trapped state change their shape by undergoing reorientation and expansion/contraction of their facets in order to decrease energy associated with them as is the case with soft particles of microgel paste.[55] This process, therefore, leads to a progressive decrease of the free energy of the whole system as a function of time. However, owing to the structural arrest, acquisition of the equilibrium structure is kinetically constrained, and therefore, the collective structural rearrangement becomes extremely sluggish demonstrating all the characteristic features of physical aging observed in the out of thermodynamic equilibrium soft materials. This physical aging, however, may not cause enhancement in elastic modulus $(G')$ as measured in the small amplitude oscillatory strain experiments and manifest itself only in the evolution of relaxation time as is the case with soft glassy materials such as microgel paste[55] and hard sphere colloidal glasses.[69,77]

Remarkably the soft glassy rheology model,[52] which describes a generic framework for aging and rejuvenation in soft glassy materials shows an excellent prediction of the rheological behavior of Carbopol dispersion.[35] The SGR model also predicts linear dependence of relaxation time on age and absence of time dependent evolution of elastic modulus as well as the yield stress,[52,78] as shown by the Carbopol dispersion in the present work. Furthermore, a structural kinetic model based on free energy minimization[79] also predicts all the rheological characteristic features of the Carbopol dispersion for $\mu=1$ while most rheological characteristic features of the Carbopol dispersion have been reported for $n=1$.[80] The fact that many predictions of the SGR theory[52,78] and the structural kinetic model[79,80] work well for the Carbopol dispersion renders insight into mechanistic aspects of aging in the same.

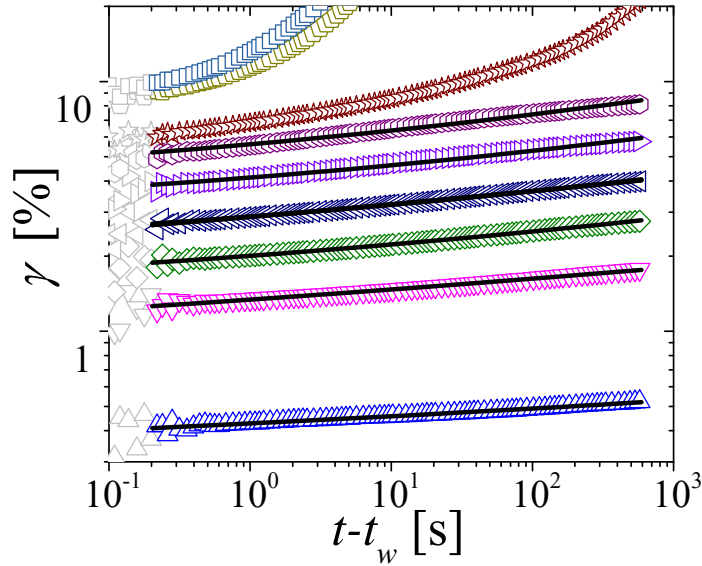

Fig 13. Evolution of strain is plotted as a function of creep time under application of constant stresses (from bottom to top 6, 9, 12, 15, 18, 21, 24, 27, 30 Pa) at the waiting time of $t_w = 600$ s. The black lines correspond to power law fits obtained by 3 parameter estimation leading to $\alpha=$ 0.06, 0.07, 0.10, 0.11, 0.12, 0.13 for increasing $\sigma$ (6, 9, 12, 15, 18, 21 Pa) respectively.

Andrade-like creep behavior

As discussed in the Introduction section, Lidon and coworkers[18] report Andrade-like creep behavior given by:

$$\gamma = \gamma_1 + \left(t/t_0\right)^\alpha, \tag{11}$$

for Carbopol dispersion over a range of stresses. Andrade (Power law) type of creep behavior leads to shear rate given by: $\dot{\gamma} = \alpha t_0^{-\alpha} t^{\alpha-1}$, suggesting that for $\alpha < 1$ the shear rate must tend to zero in a limit of large time. This behavior indicates that Andrade-like creep behavior with $\alpha < 1$ is observed only for $\sigma < \sigma_y$. For $\sigma \geq \sigma_y$, the flow must eventually set in a material leading to $\alpha \geq 1$.

We first examine the presence of Andrade-like creep behavior in the uncorrected creep curves shown in figure 5(a). A fit of the equation (11) to the experimental data is shown in figure 13. It can be seen that equation (11) shows a good fit to the data and the power law coefficient comes out to be in a range of $0.06 < \alpha < 0.13$ over the range of stresses ($\alpha$ increases as yield stress is approached). Although fit to the equation (11) is observed to be good, the power law coefficient does not show expected Andrade value of $0.3 < \alpha < 0.4$. We now analyze the behavior of the corrected creep curves given by $\bar{\gamma}$. As shown in figure 6, the creep curves of the Carbopol system used in the present work follow equation (8), given by $\bar{\gamma} = \bar{\gamma}(\xi)$. Through equation (10), the dependence of $\bar{\gamma}$ on both $t$ as well as $t_w$ is given by: $\bar{\gamma} = \bar{\gamma}\left(t^{1-\mu} - t_w^{1-\mu}\right)$. For $\mu > 1$ and in a limit of $t \gg t_w$ strain reaches a constant value.[57] On the other hand, as is the case with Carbopol suspension used in the present work, for $\mu < 1$ strain varies as: $\bar{\gamma} \approx \bar{\gamma}\left(t^{1-\mu}\right)$ in a limit of $t \gg t_w$. For an aging system, the Andrade-like behavior obtained for a single mode Maxwell model (For a single mode Maxwell model with constant modulus and dependence of relaxation time on time given by equation (9)), strain in creep flow for $t \gg t_w$ is given by: $\gamma = \gamma_0 + \left(t/\tau(t)\right) = \gamma_0 + \left(t/A'\tau_m\right)^{1-\mu}$, where $A' = A^{1/(1-\mu)}$ and $\gamma_0 = \gamma_1$.[4] Therefore, Carbopol suspension studied in the present work shows Andrade-like behavior as a special case as discussed above with $\alpha = 1 - \mu$ only in a limit of $t >> t_w$. Furthermore, since $\mu$ for the present system varies from $\mu \approx 1$ for $\sigma << \sigma_y$ to $\mu \to 0$ for $\sigma \to \sigma_y$, $\alpha$ changes from 0 to 1 with an increase in stress.

## IV. Concluding remarks and open questions

In this work, we investigate whether the aqueous suspension of Carbopol 940 undergoes physical aging. The studied Carbopol dispersion that has yield stress around 20 Pa is observed to show a very weak enhancement in the elastic modulus as a function of time. Also, in the shear start-up experiments, a peak is observed in shear stress whose magnitude increases with time elapsed since stopping the shear melting (waiting time). Both these observations suggest a possibility of aging under quiescent conditions. In the creep experiments performed at different waiting times; however, strain induced in the material shows a significant dependence on waiting time. Considering that at different waiting times, but under zero stress condition, the material shows strain recovery, we use the strain over and above the recovered strain in the analysis as necessitated by linear viscoelasticity. Interestingly this corrected strain, which shows significant waiting time dependence, shows an excellent superposition when plotted in the effective time domain, wherein real time is normalized by time-dependent relaxation time that assumes a power law dependence. Very interestingly, the power law coefficient, which shows value close to unity in a limit of small stresses decreases to zero as stress approaches the yield stress. On the other hand, over a range of stresses, not close to that of yield stress, the creep curves also show time – stress superposition suggesting the shape of the relaxation time spectrum is invariant of the aging time as well as the stress over a certain range. This work, therefore, proposes that although modulus shows only a weak change as a function of time, the creep behavior suggests a strong increase in relaxation time indicating the presence of physical aging in aqueous Carbopol suspension. We also investigate the presence of Andrade type of creep response in the Carbopol dispersion and observe that while an induced strain can be approximated to show power-law dependence on time, the power law exponent of the present system does not show unique power law coefficient.

There indeed are various issues that require further investigation. The first is work by Dinkgreve et al.,[17] which suggests that the Carbopol rheology and specifically its aging characteristics strongly depend on the preparation protocol. Particularly they do not observe thixotropic effects in gently stirred dispersion, while the present work very clearly shows physical aging in the gently stirred system. We feel that the nature of Carbopol may have an important role to play as the crosslinking density of the polyacrylic acid and therefore the softness of the particle, and size distribution of the commercially produced Carbopol particles differs from one grade to the other. It would, therefore, be interesting to establish a relation between the extent of crosslinking density, the size distribution of the particles, and the preparation protocol and the consequent absence or presence of physical aging. The present system is also expected to have a very large particle size making it an athermal system. This raises an important question whether the thermal nature of the particles a necessary precondition to induce the physical aging. If not, then whether the relaxation of the greater free energy acquired by the individual soft particles that have deformed away from their equilibrium shape during the shear melting is sufficient to induce physical aging.

Furthermore, in a review paper on yield stress fluids, Bonn et al.[3] distinguishes yield stress fluid into two categories. The first category is simple yield stress fluids, wherein shear stress depends only on the imposed shear rate. The second category is thixotropic yield stress fluids, wherein viscosity and yield stress depend on the shear history of the sample. Furthermore, the thixotropic yield stress fluids have been proposed to possess a critical shear rate below which flow does not remain stable.[17] This condition points to the presence of negative slope in stress – strain rate curve. However, in the present case of Carbopol dispersion, there is no evidence that the shear stress – shear rate curve is non-monotonic. Furthermore, yield stress remains constant and is independent of the shear history. If one goes by prediction of the simple fluidity models of Coussot and coworkers[80] (also used by Dinkgreve et al.[17]), this case corresponds to $n=1$. If one considers the free energy model of Joshi[79], this corresponds to constant modulus and $\mu=1$. Both these models, as well as the SGR model, suggests steady state stress depend on the imposed value of shear rate. This indicates that the Carbopol dispersion fulfills criterion given by a simple yield stress fluids. Furthermore, since the Carbopol dispersion's yield stress remains constant and flow curve is monotonic, it certainly violates to be categorized as "thixotropic yield stress fluid" as defined by Bonn and coworkers[3,17] However the fact still remains that Carbopol dispersion does show aging and rejuvenation. Therefore, the open question is: can this system be considered to belong to the third category as thixotropic constant yield stress system that does not show any non-monotonic flow curve. Finally, although generic models such as SGR theory[52,78] and empirical theories such as structural kinetic models[79,80] do qualitatively predict the aging behavior and rheology of Carbopol dispersion, there indeed is a need for developing a model based on Carbopol microstructure proposed in the present work.

**Acknowledgment**: We thank Profs. Gareth McKinley, Thibaut Divoux and Sebastien Manneville for helpful discussions. We acknowledge financial support from Science and Engineering Research Board, Government of India.